\documentclass[floatfix, aip, apl, preprint]{revtex4-1}

\usepackage{graphicx}
\usepackage{dcolumn}
\usepackage{bm}

\usepackage[utf8]{inputenc}
\usepackage[T1]{fontenc}
\usepackage{mathptmx}
\usepackage{etoolbox}

\usepackage{hyperref}
\hypersetup{
    colorlinks=true,
    linkcolor=blue,
    citecolor= blue,
    filecolor=magenta,      
    urlcolor=cyan,
    pdftitle={Overleaf Example},
    pdfpagemode=FullScreen,
}
\usepackage{url}

\usepackage{svg} 
\usepackage{csquotes} 
\usepackage{amsmath}
\usepackage{amsfonts}
\usepackage{amsthm}
\usepackage{amssymb}
\usepackage{siunitx}
\usepackage[english]{babel}

\usepackage{tikz}
\usepackage{pgfplots}
\usetikzlibrary{shapes.geometric, arrows, arrows.meta}
\usetikzlibrary{decorations.pathmorphing,backgrounds,positioning,}
\usetikzlibrary{external}
\usepgfplotslibrary{groupplots}
\usepgfplotslibrary{statistics}
\usepgfplotslibrary{fillbetween}

\definecolor{pink_600}{HTML}{db2777}
\definecolor{pink_500}{HTML}{ec4899}
\definecolor{pink_050}{HTML}{fdf2f8}

\definecolor{slate_050}{HTML}{f8fafc}
\definecolor{slate_100}{HTML}{f1f5f9}
\definecolor{slate_200}{HTML}{e2e8f0}
\definecolor{slate_300}{HTML}{cbd5e1}
\definecolor{slate_400}{HTML}{94a3b8}
\definecolor{slate_500}{HTML}{64748b}
\definecolor{slate_600}{HTML}{475569}
\definecolor{slate_700}{HTML}{334155}
\definecolor{slate_800}{HTML}{1e293b}
\definecolor{slate_900}{HTML}{0f172a}
\definecolor{slate_950}{HTML}{020617}

\definecolor{gray_050}{HTML}{f9fafb}
\definecolor{gray_100}{HTML}{f3f4f6}
\definecolor{gray_200}{HTML}{e5e7eb}
\definecolor{gray_300}{HTML}{d1d5db}
\definecolor{gray_400}{HTML}{9ca3af}
\definecolor{gray_500}{HTML}{6b7280}
\definecolor{gray_600}{HTML}{4b5563}
\definecolor{gray_700}{HTML}{374151}
\definecolor{gray_800}{HTML}{1f2937}
\definecolor{gray_900}{HTML}{111827}
\definecolor{gray_950}{HTML}{030712}

\definecolor{zinc_050}{HTML}{fafafa}
\definecolor{zinc_100}{HTML}{f4f4f5}
\definecolor{zinc_200}{HTML}{e4e4e7}
\definecolor{zinc_300}{HTML}{d4d4d8}
\definecolor{zinc_400}{HTML}{a1a1aa}
\definecolor{zinc_500}{HTML}{71717a}
\definecolor{zinc_600}{HTML}{52525b}
\definecolor{zinc_700}{HTML}{3f3f46}
\definecolor{zinc_800}{HTML}{27272a}
\definecolor{zinc_900}{HTML}{18181b}
\definecolor{zinc_950}{HTML}{09090b}

\definecolor{neutral_050}{HTML}{fafafa}
\definecolor{neutral_100}{HTML}{f5f5f5}
\definecolor{neutral_200}{HTML}{e5e5e5}
\definecolor{neutral_300}{HTML}{d4d4d4}
\definecolor{neutral_400}{HTML}{a3a3a3}
\definecolor{neutral_500}{HTML}{737373}
\definecolor{neutral_600}{HTML}{525252}
\definecolor{neutral_700}{HTML}{404040}
\definecolor{neutral_800}{HTML}{262626}
\definecolor{neutral_900}{HTML}{171717}
\definecolor{neutral_950}{HTML}{0a0a0a}

\definecolor{stone_050}{HTML}{fafaf9}
\definecolor{stone_100}{HTML}{f5f5f4}
\definecolor{stone_200}{HTML}{e7e5e4}
\definecolor{stone_300}{HTML}{d6d3d1}
\definecolor{stone_400}{HTML}{a8a29e}
\definecolor{stone_500}{HTML}{78716c}
\definecolor{stone_600}{HTML}{57534e}
\definecolor{stone_700}{HTML}{44403c}
\definecolor{stone_800}{HTML}{292524}
\definecolor{stone_900}{HTML}{1c1917}
\definecolor{stone_950}{HTML}{0c0a09}

\definecolor{red_050}{HTML}{fef2f2}
\definecolor{red_100}{HTML}{fee2e2}
\definecolor{red_200}{HTML}{fecaca}
\definecolor{red_300}{HTML}{fca5a5}
\definecolor{red_400}{HTML}{f87171}
\definecolor{red_500}{HTML}{ef4444}
\definecolor{red_600}{HTML}{dc2626}
\definecolor{red_700}{HTML}{b91c1c}
\definecolor{red_800}{HTML}{991b1b}
\definecolor{red_900}{HTML}{7f1d1d}
\definecolor{red_950}{HTML}{450a0a}

\definecolor{orange_050}{HTML}{fff7ed}
\definecolor{orange_100}{HTML}{ffedd5}
\definecolor{orange_200}{HTML}{fed7aa}
\definecolor{orange_300}{HTML}{fdba74}
\definecolor{orange_400}{HTML}{fb923c}
\definecolor{orange_500}{HTML}{f97316}
\definecolor{orange_600}{HTML}{ea580c}
\definecolor{orange_700}{HTML}{c2410c}
\definecolor{orange_800}{HTML}{9a3412}
\definecolor{orange_900}{HTML}{7c2d12}
\definecolor{orange_950}{HTML}{431407}

\definecolor{amber_050}{HTML}{fffbeb}
\definecolor{amber_100}{HTML}{fef3c7}
\definecolor{amber_200}{HTML}{fde68a}
\definecolor{amber_300}{HTML}{fcd34d}
\definecolor{amber_400}{HTML}{fbbf24}
\definecolor{amber_500}{HTML}{f59e0b}
\definecolor{amber_600}{HTML}{d97706}
\definecolor{amber_700}{HTML}{b45309}
\definecolor{amber_800}{HTML}{92400e}
\definecolor{amber_900}{HTML}{78350f}
\definecolor{amber_950}{HTML}{451a03}

\definecolor{yellow_050}{HTML}{fefce8}
\definecolor{yellow_100}{HTML}{fef9c3}
\definecolor{yellow_200}{HTML}{fef08a}
\definecolor{yellow_300}{HTML}{fde047}
\definecolor{yellow_400}{HTML}{facc15}
\definecolor{yellow_500}{HTML}{eab308}
\definecolor{yellow_600}{HTML}{ca8a04}
\definecolor{yellow_700}{HTML}{a16207}
\definecolor{yellow_800}{HTML}{854d0e}
\definecolor{yellow_900}{HTML}{713f12}
\definecolor{yellow_950}{HTML}{422006}

\definecolor{lime_050}{HTML}{f7fee7}
\definecolor{lime_100}{HTML}{d9f99d}
\definecolor{lime_200}{HTML}{bef264}
\definecolor{lime_300}{HTML}{a3e635}
\definecolor{lime_400}{HTML}{84cc16}
\definecolor{lime_500}{HTML}{65a30d}
\definecolor{lime_600}{HTML}{4d7c0f}
\definecolor{lime_700}{HTML}{3f6212}
\definecolor{lime_800}{HTML}{365314}
\definecolor{lime_900}{HTML}{1a2e05}
\definecolor{lime_950}{HTML}{0f1e03}

\definecolor{green_050}{HTML}{f0fdf4}
\definecolor{green_100}{HTML}{dcfce7}
\definecolor{green_200}{HTML}{bbf7d0}
\definecolor{green_300}{HTML}{86efac}
\definecolor{green_400}{HTML}{4ade80}
\definecolor{green_500}{HTML}{22c55e}
\definecolor{green_600}{HTML}{16a34a}
\definecolor{green_700}{HTML}{15803d}
\definecolor{green_800}{HTML}{166534}
\definecolor{green_900}{HTML}{14532d}
\definecolor{green_950}{HTML}{052e16}

\definecolor{emerald_050}{HTML}{ecfdf5}
\definecolor{emerald_100}{HTML}{d1fae5}
\definecolor{emerald_200}{HTML}{a7f3d0}
\definecolor{emerald_300}{HTML}{6ee7b7}
\definecolor{emerald_400}{HTML}{34d399}
\definecolor{emerald_500}{HTML}{10b981}
\definecolor{emerald_600}{HTML}{059669}
\definecolor{emerald_700}{HTML}{047857}
\definecolor{emerald_800}{HTML}{065f46}
\definecolor{emerald_900}{HTML}{064e3b}
\definecolor{emerald_950}{HTML}{022c22}

\definecolor{teal_050}{HTML}{f0fdfa}
\definecolor{teal_100}{HTML}{ccfbf1}
\definecolor{teal_200}{HTML}{99f6e4}
\definecolor{teal_300}{HTML}{5eead4}
\definecolor{teal_400}{HTML}{2dd4bf}
\definecolor{teal_500}{HTML}{14b8a6}
\definecolor{teal_600}{HTML}{0d9488}
\definecolor{teal_700}{HTML}{0f766e}
\definecolor{teal_800}{HTML}{115e59}
\definecolor{teal_900}{HTML}{134e4a}
\definecolor{teal_950}{HTML}{042f2e}

\definecolor{cyan_050}{HTML}{ecfeff}
\definecolor{cyan_100}{HTML}{cffafe}
\definecolor{cyan_200}{HTML}{a5f3fc}
\definecolor{cyan_300}{HTML}{67e8f9}
\definecolor{cyan_400}{HTML}{22d3ee}
\definecolor{cyan_500}{HTML}{06b6d4}
\definecolor{cyan_600}{HTML}{0891b2}
\definecolor{cyan_700}{HTML}{0e7490}
\definecolor{cyan_800}{HTML}{155e75}
\definecolor{cyan_900}{HTML}{164e63}
\definecolor{cyan_950}{HTML}{083344}

\definecolor{sky_050}{HTML}{f0f9ff}
\definecolor{sky_100}{HTML}{e0f2fe}
\definecolor{sky_200}{HTML}{bae6fd}
\definecolor{sky_300}{HTML}{7dd3fc}
\definecolor{sky_400}{HTML}{38bdf8}
\definecolor{sky_500}{HTML}{0ea5e9}
\definecolor{sky_600}{HTML}{0284c7}
\definecolor{sky_700}{HTML}{0369a1}
\definecolor{sky_800}{HTML}{075985}
\definecolor{sky_900}{HTML}{0c4a6e}
\definecolor{sky_950}{HTML}{082f49}

\definecolor{blue_050}{HTML}{eff6ff}
\definecolor{blue_100}{HTML}{dbeafe}
\definecolor{blue_200}{HTML}{bfdbfe}
\definecolor{blue_300}{HTML}{93c5fd}
\definecolor{blue_400}{HTML}{60a5fa}
\definecolor{blue_500}{HTML}{3b82f6}
\definecolor{blue_600}{HTML}{2563eb}
\definecolor{blue_700}{HTML}{1d4ed8}
\definecolor{blue_800}{HTML}{1e40af}
\definecolor{blue_900}{HTML}{1e3a8a}
\definecolor{blue_950}{HTML}{172554}

\definecolor{indigo_050}{HTML}{eef2ff}
\definecolor{indigo_100}{HTML}{e0e7ff}
\definecolor{indigo_200}{HTML}{c7d2fe}
\definecolor{indigo_300}{HTML}{a5b4fc}
\definecolor{indigo_400}{HTML}{818cf8}
\definecolor{indigo_500}{HTML}{6366f1}
\definecolor{indigo_600}{HTML}{4f46e5}
\definecolor{indigo_700}{HTML}{4338ca}
\definecolor{indigo_800}{HTML}{3730a3}
\definecolor{indigo_900}{HTML}{312e81}
\definecolor{indigo_950}{HTML}{1e1b4b}

\definecolor{violet_050}{HTML}{f5f3ff}
\definecolor{violet_100}{HTML}{ede9fe}
\definecolor{violet_200}{HTML}{ddd6fe}
\definecolor{violet_300}{HTML}{c4b5fd}
\definecolor{violet_400}{HTML}{a78bfa}
\definecolor{violet_500}{HTML}{8b5cf6}
\definecolor{violet_600}{HTML}{7c3aed}
\definecolor{violet_700}{HTML}{6d28d9}
\definecolor{violet_800}{HTML}{5b21b6}
\definecolor{violet_900}{HTML}{4c1d95}
\definecolor{violet_950}{HTML}{2e1065}

\definecolor{purple_050}{HTML}{faf5ff}
\definecolor{purple_100}{HTML}{f3e8ff}
\definecolor{purple_200}{HTML}{e9d5ff}
\definecolor{purple_300}{HTML}{d8b4fe}
\definecolor{purple_400}{HTML}{c084fc}
\definecolor{purple_500}{HTML}{a855f7}
\definecolor{purple_600}{HTML}{9333ea}
\definecolor{purple_700}{HTML}{7e22ce}
\definecolor{purple_800}{HTML}{6b21a8}
\definecolor{purple_900}{HTML}{581c87}
\definecolor{purple_950}{HTML}{3b0764}

\definecolor{fuchsia_050}{HTML}{fdf4ff}
\definecolor{fuchsia_100}{HTML}{fae8ff}
\definecolor{fuchsia_200}{HTML}{f5d0fe}
\definecolor{fuchsia_300}{HTML}{f0abfc}
\definecolor{fuchsia_400}{HTML}{e879f9}
\definecolor{fuchsia_500}{HTML}{d946ef}
\definecolor{fuchsia_600}{HTML}{c026d3}
\definecolor{fuchsia_700}{HTML}{a21caf}
\definecolor{fuchsia_800}{HTML}{86198f}
\definecolor{fuchsia_900}{HTML}{701a75}
\definecolor{fuchsia_950}{HTML}{4a044e}

\definecolor{pink_050}{HTML}{fdf2f8}
\definecolor{pink_100}{HTML}{fce7f3}
\definecolor{pink_200}{HTML}{fbcfe8}
\definecolor{pink_300}{HTML}{f9a8d4}
\definecolor{pink_400}{HTML}{f472b6}
\definecolor{pink_500}{HTML}{ec4899}
\definecolor{pink_600}{HTML}{db2777}
\definecolor{pink_700}{HTML}{be185d}
\definecolor{pink_800}{HTML}{9d174d}
\definecolor{pink_900}{HTML}{831843}
\definecolor{pink_950}{HTML}{500724}

\definecolor{rose_050}{HTML}{fff1f2}
\definecolor{rose_100}{HTML}{ffe4e6}
\definecolor{rose_200}{HTML}{fecdd3}
\definecolor{rose_300}{HTML}{fda4af}
\definecolor{rose_400}{HTML}{fb7185}
\definecolor{rose_500}{HTML}{f43f5e}
\definecolor{rose_600}{HTML}{e11d48}
\definecolor{rose_700}{HTML}{be123c}
\definecolor{rose_800}{HTML}{9f1239}
\definecolor{rose_900}{HTML}{881337}
\definecolor{rose_950}{HTML}{4c0519}

\pgfplotscreateplotcyclelist{custom}{
    {sky_600, mark= none},
    {violet_600, mark= none},
    {pink_600, mark= none},
    {red_600, mark= none},
    {amber_600, mark= none},
    {lime_600, mark= none},
    {lime_500, mark= none},
    {yellow_500, mark= none},
    {amber_500, mark= none},
}

\pgfplotscreateplotcyclelist{custom bar}{
    {sky_600, fill= sky_200, mark=none},
    {violet_600, fill= violet_200, mark=none},
    {pink_600, fill= pink_200, mark=none},
}
\pgfplotsset{
    compat= 1.18,
    width= 0.9\textwidth,
    height= 0.35 \textheight,
    cycle list name= custom,
}

\pgfplotsset{
    every axis plot/.append style={
        line width= 0.5,
        mark size= 1pt,
    }
}

\pgfplotsset{
    title style= {font= \small, yshift= -6pt},
    legend style= {font= \small},
    label style= {font= \small, },
    ticklabel style= {font= \scriptsize},
}

\pgfplotsset{
    colormap= {reds}{rgb255(0cm)= (254,242,242) rgb255(1cm)= (69,10,10)} 
}
\pgfplotsset{
    colormap= {blues}{rgb255(0cm)= (239,246,255) rgb255(1cm)= (23,37,84)} 
}

\renewcommand{\vec}[1]{\bm{#1}}

\DeclareMathOperator{\tr}{tr}

\makeatletter
\def\@email#1#2{%
 \endgroup
 \patchcmd{\titleblock@produce}
  {\frontmatter@RRAPformat}
  {\frontmatter@RRAPformat{\produce@RRAP{*#1\href{mailto:#2}{#2}}}\frontmatter@RRAPformat}
  {}{}
}%
\makeatother
\begin{document}

\preprint{AIP/123-QED}

\title[Data Driven Programming of Photonic Integrated Circuits]{Data Driven Programming of Photonic Integrated Circuits}
\author{G. Cavicchioli}
 \altaffiliation[Also at ]{ Institut für Hochfrequenz- und Halbleiter-Systemtechnologien, Technische Universität Berlin, Berlin, Germany}
 \email{g.cavicchioli@tu-berlin.de.}
\author{G. Masini}%
\author{F. M. Sances}
\author{F. Morichetti}
\author{A. Melloni}
\affiliation{ 
Dipartimento di Elettronica, Informazione e Bioingegneria,\\Politecnico di Milano, Milano, Italy
}%

\date{\today}

\begin{abstract}
Programming photonic integrated hardware often reveals as a challenging task because of the presence of non-idealities in the photonic chip. These include fabrication imperfections and parasitic effects such as thermal crosstalk, which cause unwanted coupling between control signals. Traditional control methods based on idealized models often fail to account for these phenomana, leading to significant discrepancies between the desired and actual circuit behaviour. In this work, we propose a data-driven approach for controlling meshes of thermally tuneable Mach Zehnder interferometers (MZIs), which exploits a machine learning (ML) model trained to compensate for these non-idealities by pre-adjusting the electrical power given to integrated phase shifters. The proposed ML system is assessed using synthetic datasets and experimentally validated on a $3 \times 3$ triangular MZI mesh. Results demonstrate that the  data-driven controller significantly improves programming accuracy, offering a robust solution for accurate programming of photonic integrated circuits.
\end{abstract}

\maketitle

\section{Introduction}
%
%
%
%
Programmable photonic integrated circuits (PICs) have emerged in the last years as essential components in increasingly complex optical systems \cite{Bogaerts2020, Harris2018}. Thanks to the integration of a large number of optical components in a compact footprint \cite{Zhu2023}, these devices are today a solid alternative to bulky free-space optical setups \cite{Zhong2020}.  Integration significantly increases thermal and mechanical stability, and enables functional tuning at run-time, opening the path to new  applications in the field of imaging and microscopy \cite{RoquesCarmes2024} \cite{Buetow2022}, sensing \cite{Miller2020} and optical communications \cite{SeyedinNavadeh2024}.

\begin{figure*}
    \centering
    \includesvg[inkscapelatex= false, width= \textwidth]{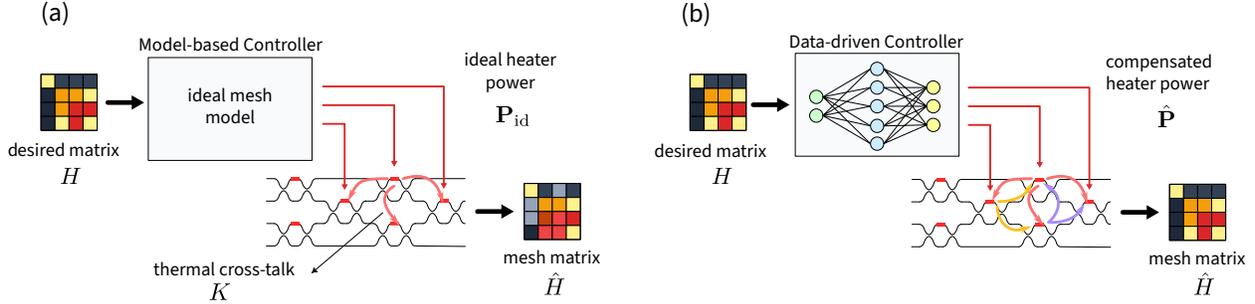}
    \caption{Conceptual picture of the proposed data-driven control model: (a) Because of cross-talk effects and other non-idealities, controlling a MZI mesh assuming a theoretically-developed model inevitably leads to programming errors. (b) Employing a Data-driven controller trained on the actual behavior of the system can compensate the non-idealities and increase the accuracy in the realized transfer matrix of the circuit.}
    \label{fig:ANN_idea}
\end{figure*}

Programmable PICs can be realized according to different architectures, including coherent and incoherent cross-bars \cite{Giamougiannis2023} \cite{Feldmann2021} and Mach-Zehnder Interferometer Meshes \cite{Perez2017}. All these architectures share the use of tuneable elements which introduce local changes in the optical length of the PIC waveguides which then reflect into a change of the properties of the circuit. These actuators can leverage different physical phenomena as electro-optic effect, acousto-optic\cite{Boyd2020} and many others, but the most common phase shifters are those based on the thermo-optic effect \cite{Liu2022}. Regardless of exploited physical effect,  fabrication errors and parasitic effects cause deviations of the actual behavior of the circuit with respect to the nominal one. As a result, the practical use of a programmable PIC without a prior calibration phase. However, the calibration of each phase shifter in a programmable PIC is not a simple task, because it requires to identify the effect of each phase shifter on the optical transfer function of the whole circuit. As schematically shown in Fig.  \ref{fig:ANN_idea}(a), this task is complicated by  the presence of cross-talk effects and other parasitic  effects, which are typically difficult to model in a realistic PIC simulations\cite{DeRose2014} 

In the last years several approaches have been proposed to solve the problem of accurately controlling programmable PICs. These can roughly be divided into two categories: \emph{open-loop approaches}, where the control of the PIC relies on a previous calibration and generation of look-up-tables (LUTs) \cite{Milanizadeh2020} \cite{Alexiev2021} or \emph{closed-loop approaches}, where the working point of the PIC is monitored through internal or external photodetectors and driving signal applied to the phase shifters is optimized in real-time with a feedback-loop control \cite{Zanetto2021} \cite{Tria2025}. Self-configuring algorithms used in Refs. \cite{Pai2020} \cite{Miller2013} belong to the latter approach. 

Yet, none of the previous approaches have proven to be suitable for every scenario. As an example, the use of LUTs is effective when the circuit is required to operate in a finite and restricted number of states, as in the case of a switch matrix, but it is not as effective when the PIC response must be continuously tuneable as in the case of MZI meshes used for matrix-vector-multiplication (MVM) \cite{Zhou2022}. On the other hand, approaches based on the isolation and calibration of each phase shifter are dependent on the specific architecture employed. 

In this work we present a control system for programmable MZI meshes, which is based on a deep-learning (DL) model. The model is trained to predict the correct set of signals (voltages or electrical powers) to set the phase shifters of the MZI mesh and obtain the desired optical transfer function  (Fig. \ref{fig:ANN_idea}(b)). Similar approaches have been employed so far only in the creation of surrogate models to be used in closed-loop control systems \cite{Cem2023a} \cite{Cem2023b}, and not in the development of a data-driven control system aimed at programming a PIC. Notably, the proposed DL model does not contain any prior knowledge of the mesh architecture making it a flexible solution which can be applied to different circuit architecture. 

The material is organized as follows: we start from a precise formulation of the control task that the DL system is required to solve (Sec. \ref{sec:ANN_problem}), followed by a description of the model employed to solve the task (Sec. \ref{sec:ANN_sol}).  The performance of the proposed model are first evaluated using synthetic datasets obtained through an MZI mesh simulator. Specifically, we focus on how the properties of the training dataset affect the model's predictions  (Sec. \ref{sec:ANN_ML_design}) and we present numerical simulation validating our approach  (Sec. \ref{sec:ANN_Num_Results}). Finally, we apply this data-driven approach to the control of a real $3 \times 3$ triangular mesh and demonstrate experimentally that the proposed controller can effectively program a mesh of MZIs (Sec. \ref{sec:ANN_exp}) and we discuss potential extensions of this approach in Sec. \ref{sec:discussion}. 

\section{Problem Formulation}
\label{sec:ANN_problem}
\noindent A MZI mesh is a multi-port photonic circuit characterized by an $N \times N$ complex transfer function $H$, where $N$ is the number of I/O optical waveguides \cite{Harris2018}. Each MZI performs a $2 \times 2$ unitary transformation. The specific expression of $H$ depend on the phase unbalance introduced in each MZI, and, depending on the number of phase shifters in each MZI, $H$ can span across different classes of transformations. As shown in Fig. \ref{fig:ANN_idea}, in this work we assume that each MZI has one phase shifter in the inner arms of the interferometer, which is used to control the amplitude response of the mesh. Controlling the phase response of the MZI mesh would require one additional phase shifter at the input of each interferometer, and the extension of the proposed technique to this case is discussed in Sec. \ref{sec:discussion}. 

Because of thermal cross-talk or other parasitic couplings, a phase shifter can induce unwanted phase shifts in neighboring waveguides in addition to its target waveguide. In the case of thermo-optic effect, this phenomenon is linear, so that the phase shift $\theta_i$ on the $i-th$ waveguide can be written as a linear combination of the electrical powers $\vec{P} = (P_i)$ fed to all the phase shifters of the circuit. Formally, the vector of phase shifts $\vec{\theta}$ can then be written as
\begin{equation}
    \vec{\theta} = K \vec{P},
\end{equation}
where $K$ is known as the \emph{cross-talk matrix}. Depending on the mesh topology and on the distance from the considered waveguide, some elements of the $K$ matrix will be more significant, while other elements will be negligible. Some techniques, such as the \emph{thermal eigenvalue decomposition} (TED) \cite{Milanizadeh2019}, have been developed to compensate cross-talk effects and achieve an accurate programming of programmable PICs; however, their implementation requires knowledge of the elements of $K$, whose estimation is complex in practice. Indeed, the elements of $K$ depend on the thermal resistance between a phase shifter and a waveguide which depend on the layout of several circuit layer including the metallic ones \cite{DeRose2014}.

For this reason, setting the heaters' powers $\vec{P}$ that realize a desired transfer matrix $H$ is a challenging task. Moreover, it requires the solution of an ill-posed \emph{inverse problem}: given a desired transfer matrix $H$, we need to find the correct values for each element in $\vec{P}$, but function $f(\cdot)$ connecting $\vec{P}$ to $H$ is nonlinear and periodic. Traditionally, inverse problems have always been difficult to solve analytically and in practice their solution often relies on optimizing the input of the forward problem until the desired output is achieved, an approach conceptually equivalent to a closed-loop control system. In the last years however deep learning was proven as a promising alternative for solving this kind of problems in various contexts including imaging\cite{Park2023} \cite{Lundervold2019}, control theory \cite{Zhao2024}, and even the design of nanophotonic circuits \cite{Molesky2018} \cite{Jiang2021}. Our approach solves the problem of programming a MZI mesh by using a discriminative artificial neural network (ANN) trained to approximate the inverse relationship  $f^{-1}(\cdot)$, returning a set of electrical powers $\vec{P}$ which  implement the target matrix $H$.

\section{Deep Learning Model for MZI Mesh Programming}
\label{sec:ANN_sol}

\begin{figure}[tb]
    \centering
    \includesvg[inkscapelatex= false, width= 1\columnwidth]{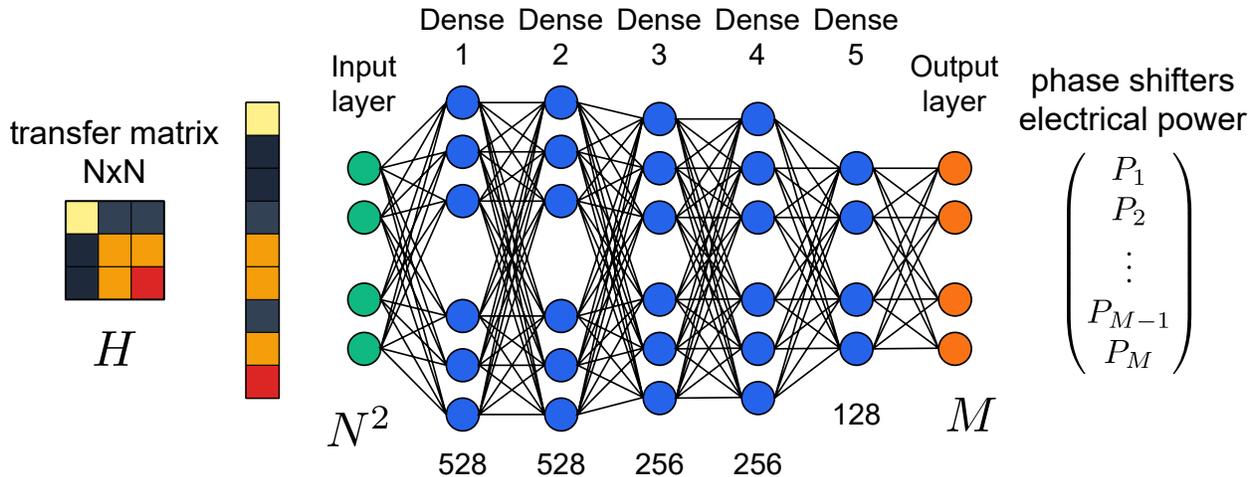}
    \caption{Architecture of the ANN used to model the relationship between the transfer matrix of a $N\times N$ MZI mesh and its $M$ phase shifters. The network takes as input the elements of the transfer matrix flattened into a vector with $N^2$ elements and returns a set of vectors representing the optimal control variables $\hat{\vec{P}}$ to realize the input transfer matrix $H$.}
    \label{fig:ANN_architecture}
\end{figure}

\noindent For the implementation of the proposed model we considered a fully connected (FC) ANN consisting of five dense layers with a decreasing number of neurons per layer. The structure of the ANN is illustrated in Fig.~\ref{fig:ANN_architecture}. The neural network takes as input the elements of a $N \times N$ complex transfer matrix, separated into real and imaginary parts and arranged into a vector of $2N^2$ real-valued elements; the output of the network is a real vector of $M$ elements, one for each phase shifter in the mesh, representing the vector of electrical powers $\vec{P}$ to be fed to the heaters.

\begin{figure*}
    \centering
    \includesvg[pretex= \tiny, width= \textwidth]{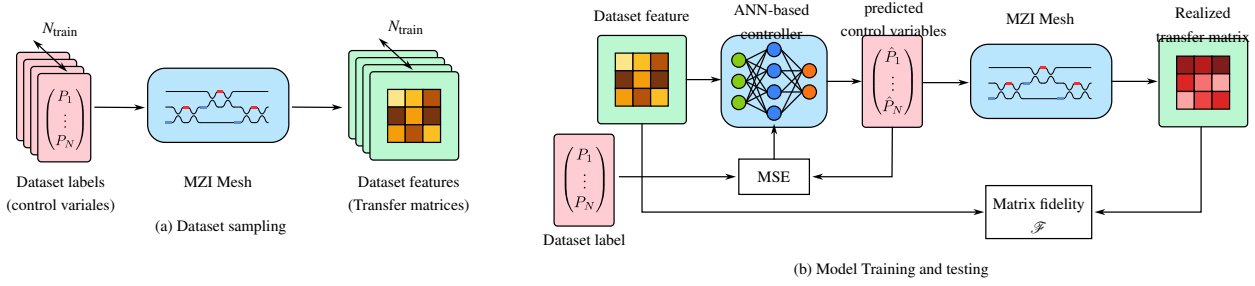}
    \caption{(a) Schematic representation of the protocol for acquiring a training dataset of $N_\text{train}$ records: a set of control variables $\vec{P}$, i.e. the labels of the dataset, is used to probe the MZI mesh, and for each target the corresponding transfer matrix, which represents the dataset features, is sampled. (b) The ANN model is trained by minimizing the $MSE$ between the predicted control variables $\hat{\vec{P}}$ and the labels of the dataset $\vec{P}$. During the testing phase, instead, $\hat{\vec{P}}$ is fed to the MZI mesh and the fidelity between the target transfer matrix $H$ and the realized one $\hat H$ is evaluated.}
    \label{fig:ANN_training_schematic}
\end{figure*}

This model can be trained using a supervised learning approach. In this formulation the transfer matrices and the electrical power fed to the heaters are respectively the \emph{features} and the \emph{labels} of the  training dataset, which consists of tuples $(H_i, \vec{P}_i)$ obtained solving the forward problem, as schematically shown in  Fig.~\ref{fig:ANN_training_schematic}.(a). The training of the ANN, illustrated in Fig.~\ref{fig:ANN_training_schematic}.(b), is performed by minimizing the cost function $\mathcal{L}$, defined as the mean squared error (MSE) between the heater powers predicted by the network, $\hat{\vec{P}}(H_i)$, and the corresponding labels $\vec{P}$ in the dataset, that is
\begin{equation}
    \mathcal{L} = MSE(\hat{\vec{P}}(H_i), \vec{P}) = \sum_{j= 1}^{M} |\hat{P}_j(H_i) - P_j|^2.
\end{equation}
For training the ANN model, the dataset was split into three subsets: \emph{training, validation, and testing datasets}. The Adam optimizer \cite{Kingma2017} was employed to minimize the cost function. To enhance the convergence of the optimization algorithm, dropout was applied to neurons in all layers, effectively acting as a form of loss function regularization \cite{Srivastava2014}. Additionally, early stopping with a patience criterion was implemented to mitigate potential overfitting \cite{Goodfellow2016}.

The performance of the model is evaluated on the test dataset using $\mathcal{L}$ as a figure of merit (FOM). Specifically $\mathcal{L}$ measures how close the control variables $\hat{\vec{P}}(H_i)$ predicted by the model are to the control variable $\vec{P}$ used to that specific transfer matrix. However, the quality of the system is primarily measured on $H$ and not on $\vec{P}$, so a better FOM is the fidelity $\mathcal{F}$ between the requested target matrix $H$ used as input of the ANN and $\hat{H}$ defined as
\begin{equation}
    \mathcal{F} = \frac{\left| \tr\left(\hat{H}H^\dagger\right) \right|}{N}.
\end{equation}
Since we are dealing with MZI meshes whose transfer matrix is unitary, we have $\mathcal{F} \in [0, 1]$ and $\mathcal{F} = 1$, as long as $\hat{H} = H$. For convenience, we also introduce the fidelity error $\mathcal{E} = 1 - \mathcal{F}$.

\section{Training Dataset Design}
\label{sec:ANN_ML_design}

\noindent While developing a deep learning model, it is essential to focus not only on refining the model itself but also on ensuring that a suitable training dataset is available. Specifically, the dataset must contain sufficient information to accurately represent the relationship the model aims to learn. To achieve this, before collecting experimental data from a physical device, we calibrated its characteristics using a custom circuit simulator of MZI meshes as a surrogate model for a real device. The simulator computes the mesh transfer matrix by multiplying  the $2 \times 2$ transfer matrix of each MZI \cite{Pai2020} and accounts for the presence of mutual cross-talk effects among the phase shifters. In the simulator $\vec{P}$ is constrained in order to introduce phase shifts $\vec{\theta}$ in the range $\left[0, \pi\right]$, such that the periodicity in the MZI response is neglected.
In our simulation, we considered a \qty{10}{\percent} cross-talk between first-neighbors phase shifters and \qty{5}{\percent} between second-neighbors ones.

A synthetic dataset composed by two parts was employed:
\begin{itemize}
    \item a sub-dataset $D_\text{rand}$ obtained by randomly sampling the control variables from independent uniform distributions,
    \item a sub-dataset $D_\text{sweep}$  by systematically sweeping a single actuator at a time, while keeping the value of all the others fixed.
\end{itemize}
More specifically, $D_\text{sweep}$ is obtained by varying a selected actuator $P_j$ across $M$ linearly spaced samples within the interval $\left[0, \pi\right]$ while the values of all the other phase shifters is held constant. This linear sweep is performed several times for each phase shifter, selecting every time a different random configuration for the fixed phase shifters. The purpose of $D_\text{rand}$ is to represent the entire state space of possible transfer matrices, while $D_\text{sweep}$ ensures that the sinusoidal response of each actuator is well captured in the final dataset.

To make the simulation more realistic, the synthetic datasets also accounted for the presence of noise both in the actuation of phase shifters and in the measurement of transfer matrices. Actuation noise was modeled by adding Gaussian noise $n_{\vec{P}} \sim \mathcal{N}(0, \sigma_P)$ to the phase shifts in the dataset before computing the associated transfer matrix. Similarly, reading noise affecting the transfer matrix was introduced as additive Gaussian noise $n_H \sim \mathcal{N}(0, \sigma_H)$, assuming for simplicity $\sigma_H = \sigma_{\vec{P}} = 2^{ENOB}$, where $ENOB$ represents the \emph{effective number of bits} of the simulated measurement system.

The dataset design parameters considered in this work are:
\begin{itemize}
    \item $N_\text{rand}$ and $N_\text{sweep}$, controlling the size and the composition of the dataset;
    \item the $ENOB$ determining the level of the noise affecting the dataset entries.
\end{itemize}

\begin{figure}
    \centering
    \includegraphics[width = 0.6\textwidth]{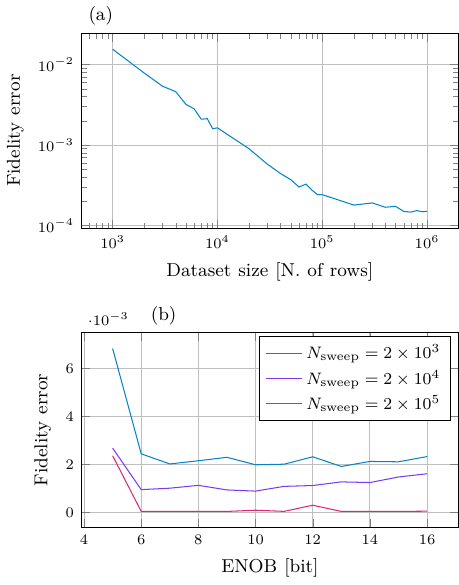}
    \caption{Fidelity error as a function of (a) the random dataset size $N_\text{rand}$ and (b) the effective number of bits (ENOB) for $N_\text{rand}$ and different size of the sweep dataset $N_\text{sweep}$.}
    \label{fig:ANN_dat_design}
\end{figure}

We evaluated the dataset necessary for training a $3 \times 3$ MZI mesh, which requires controlling three MZIs and, consequently, three phase shifts. Figure \ref{fig:ANN_dat_design} illustrates the performance of this training in terms of fidelity error $\mathcal{E}$. As shown in Fig. \ref{fig:ANN_dat_design}(a), the fidelity error $\mathcal{E}$ decreases as $N_\text{rand}$ increases. This outcome is expected since a larger dataset generally provides a more comprehensive representation of the phenomenon the neural network is designed to learn. The average fidelity error remains around \qty{1}{\percent} even for a relatively small dataset with \num{1e3} entries and appears to stabilize when the dataset size reaches $N_\text{rand} \geq \num{1e5}$. This suggests that the information the neural network can extract from the dataset is saturated, either because its learning capacity has been fully utilized or because the dataset effectively captures the complete behavior of the mesh, making additional data redundant. Nevertheless, the fidelity error stabilizes slightly above \num{1e-4}, which is well within the acceptable range for typical MZI mesh applications.  It should be noted that acquiring a dataset with more than $10^5$ entries is feasible, as thermal actuators can operate at frequencies of up to some tens of \unit{\kilo\hertz}. Even with a conservative sampling rate of \qty{10}{\kilo\hertz}, a dataset with $N_\text{rand} = \num{1e6}$, can be collected in less than five minutes.

The effect of noise is evident in Fig.~\ref{fig:ANN_dat_design}.(b), which illustrates the training results for datasets with a fixed $N_\text{rand} = 10^5$ and varying $N_\text{sweep}$, evaluated across different ENOB values. Notably, the fidelity error remains unaffected as long as $ENOB \geq 6$. Conversely, increasing $N_\text{sweep}$ consistently reduces $\mathcal{E}$, regardless of the noise level. These results indicate that the system is robust to the presence of low-to-moderate noise and that using low-noise instrumentation is not mandatory, making the analog-front end of the control system easier to design and economically more sustainable.

From this analysis of dataset properties, we then conclude that $D_\text{rand}$ plays a more significant role than $D_\text{sweep}$ in training effectiveness. Furthermore, a dataset comprising approximately \num{e6} records is sufficient to provide an accurate characterization of the circuit behavior.

\section{Numerical Results}
\label{sec:ANN_Num_Results}
\noindent After determining the optimal properties for the training dataset and finalizing the structure of the ANN, we assessed the scalability of the data-driven controller, focusing on its ability to control MZI meshes with an increasing number of inputs. More precisely, we examined triangular meshes of dimension $3 \times 3$, $4 \times 4$, and $5 \times 5$. For each configuration, we generated a synthetic dataset with identical $N_\text{rand}$ and $N_\text{sweep}$ values, trained the same ANN model depicted in Fig.~\ref{fig:ANN_architecture}, and evaluated its performance in each case.

\begin{figure}
    \centering
    \includegraphics[width = 0.7\textwidth]{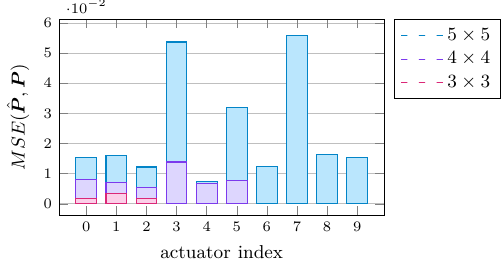}
    \caption{MSE between the predicted control variables $\hat{\vec{P}}$ and those present in the test dataset $\vec{P}$ for different triangular meshes of increasing size.}
    \label{fig:ANN_MSE_scalability}
    \bigskip
    \includegraphics[width = 0.7\textwidth]{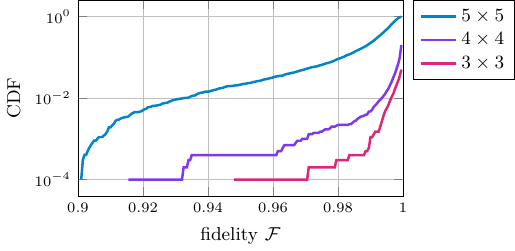}
    \caption{Cumulative distribution function (CDF) of the fidelity $\mathcal{F}$ obtained on the test dataset for triangular meshes of increasing sizes.}
    \label{fig:ANN_fid_err_scalability}
\end{figure}

In Fig. \ref{fig:ANN_MSE_scalability}, the MSE between the control variables (i.e. electrical powers applied to the phase shifters) from the test dataset $\vec{P}$ and the predicted values $\hat{\vec{P}}$ is shown. We observe that the MSE increases as the mesh size grows. This trend is expected, as the complexity of the mesh scales with the number of input parameters. The impact on the system performace is shown in Fig. \ref{fig:ANN_fid_err_scalability} which presents the cumulative distribution function (CDF) of the fidelity $\mathcal{F}$ computed over the test dataset for the three considered mesh configurations. Although performance deteriorates with increasing mesh size, the ANN remains highly accurate in its predictions, since in \qty{99}{\percent} of the cases $\mathcal{F} >$ \qty{93}{\percent}. These results demonstrate that a data-driven control system can successfully program MZI meshes of increasing increasing complexity without redesigning the ANN or tuning again the dataset properties.

\begin{figure}
    \centering
    \includegraphics[width = 0.7\textwidth]{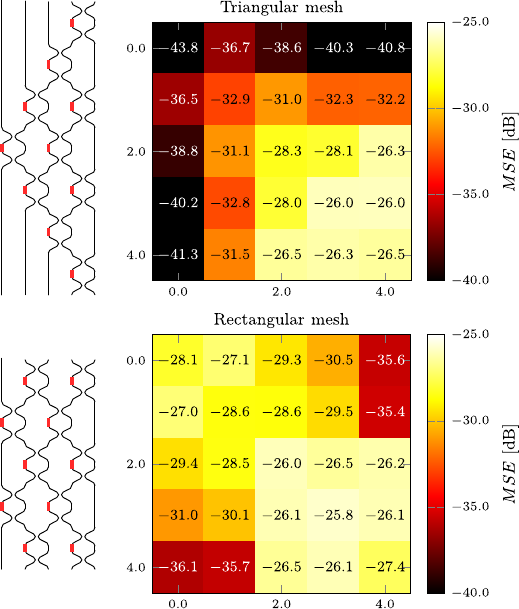}
    \caption{MSE computed element by element between the target matrix $H$ and the matrix $\hat{H}$ realized with the predicted control variables $\hat{\vec{P}}$ for a $5 \times 5$ triangular (Reck) mesh and a $5 \times 5$ rectangular (Clements) mesh.}
    \label{fig:ANN_reck_vs_clements}
\end{figure}

Finally, the performance of the model was also evaluated for different mesh architectures. Specifically, we considered two $5 \times 5$ MZI meshes with a triangular (Reck \cite{Reck1994}) and a rectangular (Clements \cite{Clements2016}) layout respectively. To assess the performance of the model, for both topologies we evaluated the $MSE$ between the elements $\hat{H}_{ij}$ of the matrix realized using the predicted control variables $\hat{\vec{P}}_{n}$ and the elements $H_{ij}$ of the matrices in the test dataset, that is 
\begin{equation}
    \label{eq:ANN_MSE_matrix}
    MSE_{ij} = \frac{1}{N_\text{test}} \sum_{n=1}^{N_\text{test}} \left|H_{ij, n} - \hat{H}_{ij}(\hat{\vec{P}}_n)\right|^2,
\end{equation}
where the index $n$ refers to the record of the test dataset.

Figure \ref{fig:ANN_reck_vs_clements} compares the average MSE for each matrix element across the two architectures. In the triangular mesh case, the neural network achieves an accuracy of at least \qty{-26}{\decibel} for all matrix elements, with significantly higher accuracy (\qty{-36}{\decibel} or better) for elements in the first row and the first column. This inhomogeneity arises from the triangular topology of the mesh. As a matter of fact, the elements in the first column and the first row are determined solely by the first diagonal and anti-diagonal row of MZI respectively. Consequently, these elements are influenced by the programming errors of fewer phase shifters, whereas elements in the lower-right portion of the transfer matrix accumulate errors from a larger number of MZIs. In a rectangular mesh instead, the average MSE is more uniformly distributed across the matrix elements while still achieving an accuracy equal or higher than \qty{-26}{\decibel}. This is expected, as in a rectangular mesh, light traveling between any input-output pair encounters, on average, the same number of MZIs. Consequently, all the paths tend to accumulate the same amount of programming errors. 

This analysis proves that our approach can be flexibly used to program MZI meshes with different topologies, removing the need to design custom control algorithms for different circuit layouts. Moreover, our approach provides an alternative to \emph{in-situ} optimization of the phase shifter value, as in the case of self-configuring algorithm, whose complexity poorly scale in the case of rectangular meshes, because of the need of a global simultaneous optimization of all the phase shifters \cite{Pai2019}. 

\begin{figure*}[t!]
    \centering
    \includesvg[pretex= \footnotesize, width= \textwidth]{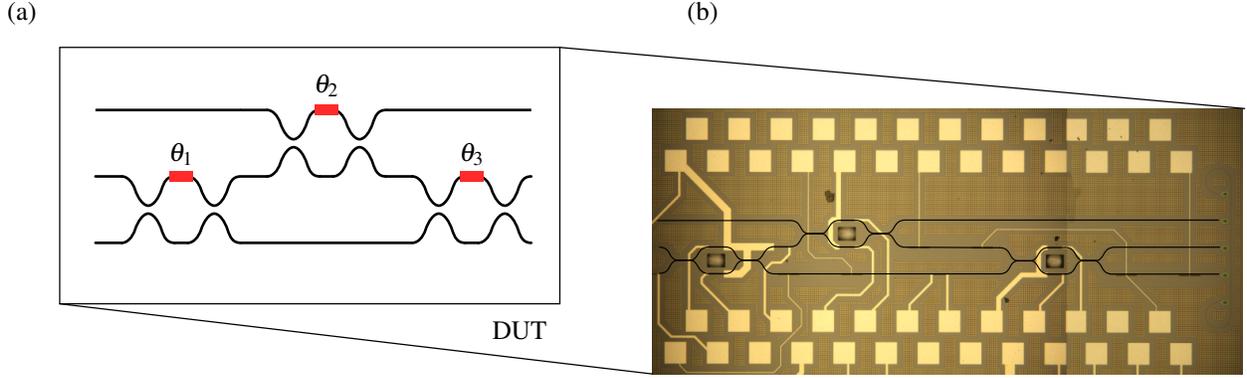}
    \caption{(a) Schematic of the $3 \times 3$ triangular mesh used as device under test (DUT) to train the neural network model. The position of the three phase shifters ($\theta_1$, $\theta_2$ and $\theta_3$) used to control the circuit is indicated. (b) Microscope photograph of the DUT. Optical waveguides are highlighted with black lines.}
    \label{fig:DUT_ANN}
\end{figure*}

\section{Experimental demonstration}
\label{sec:ANN_exp}

\noindent To validate experimentally the effectiveness and performance of the proposed ANN-based controller, we utilized a $3 \times 3$ feed-forward triangular mesh realized on a silicon photonics platform. The schematic and layout of the device under test (DUT) are shown in Fig. \ref{fig:DUT_ANN}. An experimental dataset was acquired in-situ using this circuit, which was then employed to train the controller. Once the training was completed, the trained controller was tested directly on the same DUT.

\begin{figure}
    \centering
    \includegraphics[width = 0.7\textwidth]{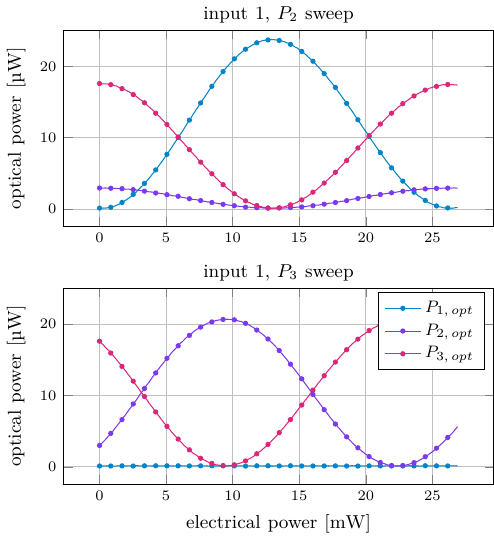}
    \caption{Characteristic of the phase shifter $P_2$ and $P_3$ observed shining the input port 1 and sweeping electrical power fed to the phase shifter with an update frequency $f_\text{samp}= \qty{25}{\kilo\hertz}$.}
    \label{fig:ANN_MZI_sweep}
\end{figure}

In the experiments, unlike in simulations, the relationship between the electrical power supplied to the thermal phase shifters $\vec{P}$, which represent our control variables, and the resulting phase shifts $\vec{\theta}$ is unknown prior to the training. However, since $H$ is periodic in $\vec{\theta}$, we are interested in phase shifts in the range $[0, \pi]$. To ensure that we apply phase shifts in this range only, before acquiring the dataset a preliminary characterization of the circuit was performed, following the same procedure for calibrating the amplitude of triangular meshes as in \cite{Pai2023a}. For each MZI, we conducted a sweep of the electrical power using $N = 256$ samples at a sampling frequency of $f_\text{samp} = \qty{25}{\kilo\hertz}$, thus ensuring that cross-talk effects remain negligible. The results of this sweep for input port 1 are shown in Fig. \ref{fig:ANN_MZI_sweep}, where the period $P_{2\pi}$ of the phase shifters $\theta_2$ and $\theta_3$ is clearly visible. The peak optical output power levels $P_\text{ref, opt}$ are used to normalize the transfer matrix elements 
\begin{equation}
    |H_{ij}|^2 = \frac{P_{ij, \; \text{opt}}}{P_{ij, \, \text{ref, opt}}}.
\end{equation}
in order to eliminate the contribution of the the input and output coupling losses.

To construct the dataset, we generated the targets $\vec{\theta}$ on a computer by sampling from a uniform distribution over the interval $[0, \pi]$. These phase shifts were then converted into electrical voltages $V_i$ applied to the phase shifters using the relationship
\begin{equation}
    V_i = R\sqrt{\frac{\theta_i}{\pi}P_\pi},
\end{equation}
where we assumed a resistance $R = \qty{330}{\ohm}$ as measured from electrical characterization of heaters. However, the actual electrical power consumed by the heaters was determined from direct measurements of the electrical current absorbed by the phase shifters, so eventual changes in the heater resistance does not affect the dataset accuracy. but only the statistical distribution of the phase shifts $\theta$.

The training dataset consists of $N_\text{rand} = 10^6$ records. For each input port of the DUT, the optical transmission was measured for various combinations of $V_1$, $V_2$, and $V_3$ with an update frequency $f_\text{write} = \qty{3}{\kilo\hertz}$, while applying an oversampling factor of 8 read samples per voltage set. This results in an effective sampling frequency $f_\text{samp} = \qty{24}{\kilo\hertz}$. These settings ensure that transient effects due to the analog electronic front-end and the slow dynamics of cross-talk are excluded. In the reported experiments, the sampling of each column of the transfer matrix requires approximately 5 minutes and 30 seconds, leading to a total sampling duration of about 17 minutes and 30 seconds. 
Regarding the ANN architecture, we retained the same layer configuration as presented in Sec. \ref{sec:ANN_sol} and Fig. \ref{fig:ANN_architecture}.

\begin{figure}[tbh]
    \centering
    \includegraphics[width = 0.8\textwidth]{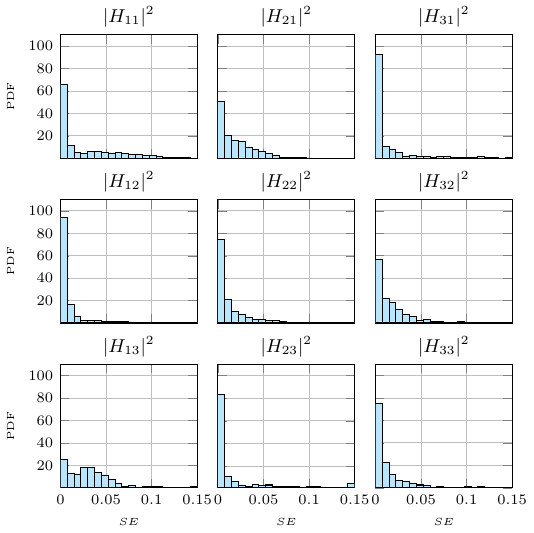}
    \caption{Probability distribution functions (PDF) of the squared error ($SE$) between the squared absolute values $|H_{ij}|^2$ of the test and of the experimentally realized transfer matrix elements.}
    \label{fig:ANN_MSE_matrix_exp}
    \bigskip
    \includegraphics[width = 0.6\textwidth]{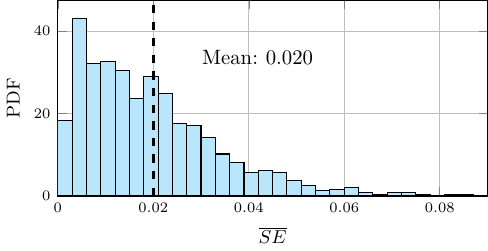}
    \caption{Probability distribution functions of the average squared error ($\overline{SE}$) between the test and the experimentally realized transfer matrix elements, averaged.}
    \label{fig:ANN_MSE_mean_exp}
\end{figure}

Finally, for testing the trained model, we used a test dataset of $N = 2000$ pairs $(\vec{P}, H)$, experimentally sampled from uniformly distributed values of $\vec{P}$. As in Sec. \ref{sec:ANN_Num_Results}, we computed the MSE between the control vectors $\vec{P}$ in the dataset and those predicted by the ANN. 
Additionally, instead of relying on a surrogate model to evaluate the transfer matrix $\hat{H}$ corresponding to $\hat{P}$, we directly applied $\hat{P}$ to the physical mesh and conducted an in-situ online testing. To evaluate the performance of the model we computed the element-wise squared error
\begin{equation}
    SE_{ij}= \left||H_{ij, n}|^2 - |\hat{H}_{ij}(\hat{\vec{P}}_n)|^2\right|^2
\end{equation}
between the absolute square of the transfer matrix elements $|H_{ij, n}|^2$ from the dataset and those experimentally inferred $|\hat{H}_{ij}(\hat{\vec{P}}_n)|^2$.
Moreover, we consider the average of the $SE$ across all the matrix elements
\begin{equation}
\overline{SE} = \frac{1}{9} \sum_{i, j} {SE_{ij}}.
\end{equation}

The model's performance is illustrated in Fig. \ref{fig:ANN_MSE_matrix_exp} and Fig. \ref{fig:ANN_MSE_mean_exp}, which present the probability density functions (PDF) of $SE_{ij}$ and $\overline{SE}$ respectively. The highest observed value for the $SE$ is approximately \qty{10}{\percent}, with an average around \qty{2}{\percent} and a PDF exhibiting an exponential decay for all elements except for $H_{31}$. This deviation is justified by the fact that the realization of this element involve the phase shifters $\theta_1$ and $\theta_2$ which belong to distinct diagonal of the mesh. Because of that, this element is influenced by higher calibration and programming errors compared to the other elements.

\begin{figure}[tbh]
    \centering
    \includegraphics[width = 0.6\textwidth]{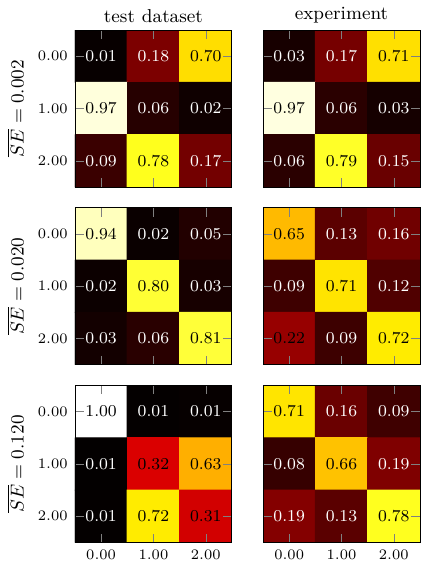}
    \caption{Comparison of the target matrix $H$ and its experimental realization $\hat{H}$, for the best, the average and the worst values of the $\overline{SE}$.}
    \label{fig:ANN_matrix_comparison}
\end{figure}

Figure \ref{fig:ANN_matrix_comparison} shows three examples of the transfer matrix realized by the controller compared to the target matrix. These selected cases  correspond to the best MSE achieved by the system ($\overline{SE}= 0.2$\%), to the average one ($\overline{SE} = 2$\%) and the worst one ($\overline{SE}= 12$\%). It must be kept into account, however, that cases with a $\overline{SE} \geq \qty{5}{\percent}$, where the realized matrix deviates significantly from the target matrix, are statistically rare, as evidenced by the probability density function in Fig. \ref{fig:ANN_MSE_mean_exp}. These results demonstrate the good performance of the proposed data-driven controller in programming MZI meshes.

\section{Conclusion}
\label{sec:discussion}
\noindent In this work we demonstrated the use of data-driven ML model to program feed-forward MZI meshes. The model does not need any previous knowledge about the PIC architecture, thus resulting in a completely black-box approach. Even in the presence of parasitic effects like thermal cross-talk, it correctly predicts the right set of driving signals for the phase actuators to achieve the desired intensity response of the circuit.

The model has been validated through numerical simulations, where its performance has been evaluated for different mesh sizes, topologies and accuracy of the training dataset. We also provided an experimental demonstration on a thermally tunable Silicon Photonic $3 \times 3$ mesh, sampling the training data in-situ and using the measured dataset to train the neural network. 
It is worth noting that the required data for the  training of the PIC can be extracted in a very short time. Thermal phase shifters allow the generation of thousands of test data every second, so that in the reported experiments the model achieves high performance with training datasets which can be sampled within a few minutes and without the need of particular low-noise electronic front-ends. Even shorter data-collection time can be achieved in PIC integrating fast electro-optic phase shifters working as sub-ns scale. Moreover, in case of wavelength independent circuits, as the one considered in this work, data acquisition could be performed in parallel for the $N$ columns of the transmission matrix by using Wavelength Division Multiplexing (WDM) schemes, thus reducing the total sampling time by a factor $N$.

The proposed approach can be extended to program also the phase response of MZI meshes. To this end,  information is required on the phase of the transfer matrix elements in the training dataset, which would require the use of phase analyzer, like coherent detectors, integrated in the photonic chip, and the extension of the ANN model to handle complex-valued data. 

\section*{Acknowledgment}

\noindent \small{\textit{This work was supported by the Italian National Recovery and Resilience Plan (NRRP) of NextGenerationEU, “Telecommunications of the Future” (PE00000001 - program “RESTART”, Structural Project “Rigoletto" and Focused Project “HePIC") and on MUSA: Multilayered Urban Sustainability Action.}}

\bibliography{bibliography}

\end{document}